\documentclass[epj]{svjour}
\input epsf

\newcounter{append}
\newcommand{\bc}{\begin{center}}
\newcommand{\ec}{\end{center}}
\newcommand{\be}{\begin{equation}}
\newcommand{\ee}{\end{equation}}
\newcommand{\ba}{\begin{array}}
\newcommand{\ea}{\end{array}}
\newcommand{\beqn}{\begin{eqnarray}}
\newcommand{\eeqn}{\end{eqnarray}}

\begin{document}

\title{Off equilibrium dynamics in 2d-XY system}

\author{St\'ephane Abriet and Dragi Karevski }
\institute{Laboratoire de Physique des Mat\'eriaux, UMR CNRS No. 7556, Universit\'e Henri
Poincar\'e (Nancy 1), B.P. 239,\\ F-54506 Vand\oe uvre l\`es Nancy cedex,
France}

\date{September 15, 2003}

\abstract
{We study the  non-equilibrium time evolution of the classical XY spin model in two dimensions.
The two-time autocorrelation and linear response functions are considered for systems initially prepared
in a high temperature state and in a completely ordered state. After a quench into the critical phase, we determine, via Monte Carlo
simulations, the time-evolution of these quantities and extract the temperature dependence of the slope of the parametric plot
susceptibility/correlation in the asymptotic regime. This slope is usually identified with the infinite fluctuation-dissipation ratio
which measures the violation to the equilibrium fluctuation-dissipation theorem. However, a direct measure of this ratio leads to a vanishing
value.
}

\authorrunning{S. Abriet and D. Karevski }
\titlerunning{Off equilibrium dynamics in 2-d XY system}


\maketitle

\maketitle

\section{Introduction}
Non-equilibrium properties of classical spin systems have received a lot of interest these last years,
especially in the context of aging\cite{cugliandolo1,crisanti} and from the point of view of
fluctuation-dissipation theorem (FDT) and its extensions.\cite{fdt}
The main feature of non-equilibrium dynamics is the breakdown of time-translation invariance,
which is a characteristic that has been used recently, together with space-symmetries in
order to build a space-time conformal like theory for some scale invariant systems.\cite{henkel1}
This theory has given some predictions that have been already tested on some systems, like the
Ising with Glauber dynamics or spherical model.\cite{henkel2}
A system relaxing towards its equilibrium state shows, in the aging regime,
a dependence in the two-time functions on both the observation time $t$ and the so called waiting
time $t_w<t$. In this context, an extension of the fluctuation-dissipation theorem (FDT) was proposed.\cite{cuglifdt}
At equilibrium, the FDT relates the correlation
function to its conjugate linear response function via
\begin{equation}
R(t-t_w)=\beta \frac{\partial}{\partial t_w}C(t-t_w)\; ,
\label{eq1}
\end{equation}
where the time enters only through the difference $t-t_w$.
Out of equilibrium, the generalisation takes the form
\begin{equation}
R(t,t_w)=X(t,t_w)\beta\frac{\partial}{\partial t_w}C(t,t_w)\; ,
\label{eq2}
\end{equation}
defining the factor $X(t,t_w)$, the so called fluctuation-dissipation ratio which measures the violation of the FDT.
It measures the ratio between
the actual response and the expected response if the FDT was valid.
Recently, a lot of interest was put in the asymptotic value of the FDT ratio, defined by
\begin{equation}
X_{\infty}=\lim_{t_w\rightarrow\infty}\lim_{t\rightarrow\infty} X(t,t_w)\; .
\end{equation}
In particular, Godr\`eche and Luck\cite{godreche} proposed that this quantity should be universal for a critical quench.
Evidences to support this universality were obtained on exactly solvable spin systems quenched from infinite temperature, numerically on 2d and 3d
Ising model with Glauber dynamics\cite{godreche2} and also checked from field-theoretic two-loop expansion of the $O(n)$ model\cite{calabrese}
and the 2d voter model.\cite{sastre}
This universality was tested recently for a wide class of initial states in 1d Glauber Ising model.\cite{henkelschutz}

However, going on more complex systems the linear response function itself is not accessible by numerical, nor experimental analyses and one is forced
to look on integrated response functions, that is measuring susceptibilities:
\begin{equation}
\chi (t,t_w)=\int_{t_w}^{t} {\rm d}t'R(t,t')\; ,
\label{eq3}
\end{equation}
where the perturbation field is applied between time $t_w$ and $t$, that is a zero-field-cooled (ZFC) scenario.
To extract information from the susceptibility on the FDT ratio, one usually plots the susceptibility versus the correlation function
and defines from the asymptotic slope of the curve, a number
\begin{equation}
X^{\chi}_{\infty}=-\lim_{C\rightarrow 0}\frac{{\rm d}\chi}{{\rm d}C}
\end{equation}
which is often identified with the FDT ratio $X_{\infty}$. 
The idea behind that is coming from the analyses of infinite-range glassy systems\cite{cugliandolo}
where asymptotically, the fluctuation-dissipation ratio depends on time only through the correlation function. One has for those systems
\begin{equation}
X(t,t_w)=X(C(t,t_w))\; .
\label{eqxc}
\end{equation}
In this context, the limiting ratio $X_{\infty}$ was interpreted as a temperature ratio between the actual inverse temperature $\beta$ and an
effective inverse temperature seen by the system:\cite{xeffectemp} $\beta_{eff}=\beta X$. From the dependence (\ref{eqxc}), 
on obtains for the susceptibility
the expression
\begin{equation}
\chi (t,t_w)=\beta\int_{C(t,t_w)}^{1} {\rm d}C' X(C')\; .
\end{equation}
This last equation strictly holds if the violation ratio is a function of $C(t,t_w)$ only and in this case, one can identify
$X_{\infty}$ with $X^{\chi}_{\infty}$.
It is in particular
the case at equilibrium since $X=1$ is a pure constant and it gives $\chi=\beta(1-C)$.

In a recent work,~\cite{chatelain} an exact expression was derived for Glauber-like dynamics which enables to
calculate directly the linear response function to an infinitesimal field. The
advantage of this approach is obvious since it gives direct access to the linear response, non-linear effects are avoided, 
and not only to the susceptibility as in the ZFC scenario.

In this context, we have performed a Monte Carlo study of the nonequilibrium evolution of the two-dimensional classical
XY system. We have studied the evolution of the system after a quench from
infinite temperature toward the low temperature critical phase, up to the Kosterlitz-Thouless point.
We have also considered the relaxation from a completely ordered initial state. For that purpose,
we have calculated two-point correlation functions, susceptibilities and response functions.
From these data, we have checked the violation of the FDT and compared
our numerics with theoretical predictions (spin wave approximation) and previous numerical works when available.
The paper is organised as follows: in the next section, we present the dynamics
of the model and its solution for two-time quantities in the spin wave approximation. The following section deals
with the numerical analyses for the ordered initial state. We turn
then to the infinite temperature initial condition. We summarise and discuss our results in the last section.

\section{Two-dimensional dynamical XY model}
The two-dimensional
ferromagnetic XY model is defined via the Hamiltonian
\begin{equation}
H=-\sum_{<ij>} S_i\cdot S_j
\end{equation}
where the sum is over nearest neighbour pairs $ij$ on a square lattice and where the classical spin variables $S_i$
are two-dimensional vector fields of unit length. Introducing angular variables, one
can rewrite the original Hamiltonian in the form
\begin{equation}
H=-\sum_{<ij>}\cos(\theta_i - \theta_j )\; .
\end{equation}
The equilibrium properties of this model are well-known since the pionneering work of Berezinskii,\cite{berezin} Kosterlitz
and Thouless and others.\cite{kosterlitz}
At a temperature $T_{KT}$, the system undergoes a continuous topological transition due to the pairing of vortex and 
anti-vortex excitations. Below the transition temperature, the system is
characterised by a line of critical points reflecting a quasi-long range ordered phase with algebraic correlation functions. 
The spin-spin correlation critical exponent $\eta$ is continuously varying with the
temperature field. Although for the spin-spin $\eta$ exponent spin-wave approximation gives an accurate analytic
prediction at low temperature,\cite{berezin} only numerical estimates are known
for the full temperature regime.\cite{berche}

The dynamics of the model was studied extensively in the context of
coarsening.\cite{bray1} The two-time spin-spin autocorrelation function and the associated linear response function have been studied
only recently in ref.\cite{berthier}. In the spin-wave approximation, valid at
low temperature($T\ll T_{KT}\simeq 0.89$), the nonconserved dynamics of the angular variable is given by the 
Langevin equation\cite{bray1}
\begin{equation}
\frac{\partial}{\partial t} \theta(x,t)=-\frac{\delta F(\theta)}{\delta \theta} +\zeta(x,t)
\end{equation}
where $\zeta(x,t)$ is a Gaussian thermal noise with variance $\langle\zeta(x,t)\zeta(x',t')\rangle=
2T\delta(x-x')\delta(t-t')$ and the free energy functional is given by\cite{bray1}
\begin{equation}
F(\theta)=\frac{\rho(T)}{2}\int {\rm d}^2 x[\nabla \theta]^2
\end{equation}
where $\rho(T)$ is the spin-wave stiffness, related to the $\eta(T)$ exponent by the relation
$2\pi\rho(T)=T/\eta(T)$.

Taking as initial condition a completely ordered state $\theta(x,0)=\theta_0$, using the previously defined spin-wave functional, it is possible to obtain analytical expressions for the two-time
autocorrelation and response functions. These reads, at enough long times, for the autocorrelation function 
$C(t,t_w)=V^{-1}\int {\rm d}^2 x \langle \cos[\theta(x,t)-\theta(x,t_w)]\rangle$:\cite{berthier}
\begin{equation}
C(t,t_w)=\frac{1}{(t-t_w)^{\eta(T)/2}}\left(\frac{(1+\lambda)^2}{4\lambda}\right)^{\eta(T)/4}\; ,
\label{sec25}
\end{equation}
where $t_w$ is the waiting time, $t$ is the total time and $\lambda=t/t_w$ is the scaling ratio.
This behaviour can be explained in the following way. At short time difference $t-t_w\ll t_w$, the fluctuations of small wavelength ($\ll \xi(t_w)$) have equilibrated and we are in a quasi-equilibrium
regime with a correlation function decaying as $C(t,t_w)\sim (t-t_w)^{-\eta(T)/z}$ where the dynamical exponent $z=2$ for the 2d XY model. At longer times, when the scaling function significantly
differs from 1, the aging process takes place giving rise to a full two-time dependence, that is a breakdown of time-translation invariance.
The conjugate response function, defined by 
$R(t,t_w)=V^{-1}\left.\int {\rm d}^2 x \frac{\delta \langle
S(x,t)\rangle}{\delta  h(x,t_w)}\right|_{h=0}$ is
\begin{equation}
R(t,t_w)=\frac{2\eta(T)}{T}\frac{C(t,t_w)}{t-t_w}\; ,
\label{sec26}
\end{equation}
with $C(t,t_w)$ given in equation~(\ref{sec25}).
It is amazing to notice at this point that the last equation has exactly the form obtained from the Fluctuation-Dissipation 
theorem with a power law equilibrium correlation function $C(t,t_w)\simeq
A(t-t_w)^{-\eta/z}$, where $z$ is the dynamical exponent. The difference from the equilibrium 
situation lies in the fact that the nonequilibrium amplitude $A$ is actually depending on time too. So, by
differentiation, one has also a term arising from the derivative of the amplitude $A(t_w)$, leading to a  deviation from the FDT.
From equations (\ref{sec25}) and (\ref{sec26}), it is easy together with the definition of the fluctuation-dissipation 
ratio given previously to obtain
\begin{equation}
X(t,t_w)=\left(1-\frac{(\lambda-1)^2}{2(1+\lambda)}\right)^{-1}\; .
\end{equation}

For a quench from an infinite temperature state to a temperature
$T<T_{KT}$, no such analytical expressions are available. 
However, on the basis of scaling arguments,\cite{bray1} one can postulate the general expressions
\begin{equation}
C(t,t_w)=\frac{1}{(t-t_w)^{\eta(T)/2}}f_C\left(\frac{\xi(t)}{\xi(t_w)}\right)
\label{sec28}
\end{equation}
and 
\begin{equation}
R(t,t_w)=\frac{1}{(t-t_w)^{1+\eta(T)/2}}f_R\left(\frac{\xi(t)}{\xi(t_w)}\right)
\label{sec29}
\end{equation}
where $f_C$ and $f_R$ are the scaling function and where the correlation length $\xi$ has a different behaviour if the quench is done from
infinite temperature or from a completely ordered initial state, namely one has:\cite{bray2}
\begin{equation}
\xi(t)\sim \left\{
\begin{array}{ll}
t^{1/2}&\qquad T_i<T_{KT}\\
(t/\ln t)^{1/2}&\qquad  T_i>T_{KT}
\end{array} \right.
\label{sec30}
\end{equation}
The logarithmic correction in the disordered initial state case is due to the slowing down of the coarsening
caused by the presence of free vortices,\cite{bray2} since the approach toward equilibrium proceeds through the annihilation of vortex-antivortex pairs,
a process which is slower than the equilibration of spin waves.

We shall concentrate first on checking numerically these analytical predictions, testing at the same time the validity of our numerics, and
then turn to the numerical study of the infinite-temperature initial condition.

\section{Numerics}
\subsection{Numerical approach}
During the simulations, the system is initially
prepared in two different ways: spin angles $\theta_i$ chosen at random in the interval $[0,2\pi]$ corresponding to the infinite temperature
initial state and constant initial angles, $\theta_i=cst.$ $\forall i$, corresponding to the zero temperature initial state.
For the numerical analysis, we use a standard metropolis dynamics where a spin chosen at random is turned at random
with an acceptance probability given by $\min[1,\exp(-\Delta E/T)]$ where $\Delta E$ is the difference energy between the actual
configuration and the former one. As stated before, in order to go beyond the susceptibility and to access directly the response itself, we use a
different dynamics, Glauber-like, where the transition probabilities of a configuration with a spin $S_i$ to a new value $S'_i$ is given by
\begin{equation}
p(S_i\rightarrow S'_i)=\frac{W(S'_i)}{W(S'_i)+W(S_i)}
\end{equation}
with
\begin{equation}
W(S_i)=\exp\left(-\frac{1}{T}S_i\sum_j S_j\right)\; .
\end{equation}
Both dynamics have the same dynamical exponents and one expect no significant changes for thermodynamical quantities.

The two-times autocorrelation function is defined by
\begin{equation}
C(t,t_w)=\frac{1}{L^2}\sum_i\langle \cos[\theta_i(t)-\theta_i(t_w)]\rangle
\end{equation}
where $\langle . \rangle$ is the average over the thermal histories.
In the metropolis simulation, we calculate the ZFC susceptibility via\cite{barrat}
\begin{equation}
\chi(t,t_w)=\frac{1}{L^2h^2}\sum_{i}\overline{\langle h_i\cdot S_i(t)\rangle}
\end{equation}
where $h$ is a small bimodal random magnetic field applied from $t_w$. The overline means an average over the field realizations. 
Practically in our simulations we use the value $h=0.04$.\cite{berthiercompare}

The response function itself is obtained numerically with the help of the Glauber-like dynamics.\cite{chatelain} 
By definition, the autoresponse to an
infinitesimal magnetic field applied at $t_w$ is given by
\begin{equation}
R(t,t_w)=\frac{\delta S_i(t)}{\delta h_i(t_w)}\; .
\end{equation}
With the help of the master equation 
\begin{equation}
P(\{\theta'\},t+1)=\sum_{\{\theta\}}p(\{\theta\}\rightarrow\{\theta'\})P(\{\theta\},t)
\end{equation}
and following the lines of ref.\cite{chatelain}, it is easy to arrive at
\begin{eqnarray}
R(t,t_w)&=&\beta\langle\cos \theta_i(t)\left[\cos\theta_i(t_w+1)-\cos\theta_i^w(t_w+1)\right]\rangle\nonumber\\
+&\beta\langle&\sin \theta_i(t)\left[\sin\theta_i(t_w+1)-\sin\theta_i^w(t_w+1)\right]\rangle
\label{resp1}
\end{eqnarray}
where $\cos\theta_i^w(t)$ and $\sin \theta_i^w(t)$ are the components of the Weiss magnetisation given by
\begin{equation}
S_i^{x,y}=\left.\frac{1}{\beta}\frac{\partial}{\partial h_{x,y}}\ln Z_i\right|_{h=0}\; .
\end{equation}
$Z_i=\exp(-\beta H(\theta_i,h))+\exp(-\beta H(\theta'_i,h))$ is the local partition function in the field.

The thermodynamical quantities are calculated on square samples with periodic boundary conditions of linear size up to $L=512$ and averaged
typically over 1000 thermal histories.

\subsection{Ordered initial state}
We start with a completely ordered state and set the temperature $T<T_{KT}$, in order first to check the compatibility 
of our numerics with the
analytical predictions in the spin-wave approximation. In figure~1, we present the results obtained for the autocorrelation function
in the asymptotic regime $t-t_w\gg t_w\gg 1$ for a temperature of $T=0.3$ where the expression~(\ref{sec25}) is expected to hold.
For different waiting times,
the collapse of the data is fairly good and the power-law behaviour in terms of the variable $(1+\lambda)^2/(4\lambda)$ gives a very good
agreement with the XY $\eta(T)$ exponent, as it can be seen on figure~2.
\begin{figure}[ht]
\epsfxsize=7cm
\begin{center}
\mbox{\epsfbox{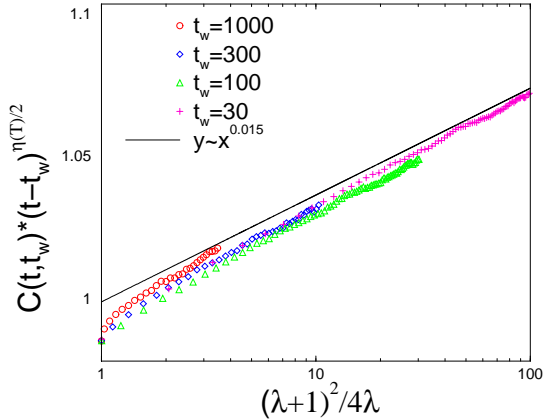}}
\end{center}
\caption{Rescaled autocorrelation function at $T=0.3$ for a system of linear size $L=512$ and for different waiting times.
The solid line is guide for the eyes corresponding to the value
$\eta(0.3)\simeq 0.015$. \label{fig1}
}
\end{figure}
\begin{figure}[ht]
\epsfxsize=7cm
\begin{center}
\mbox{\epsfbox{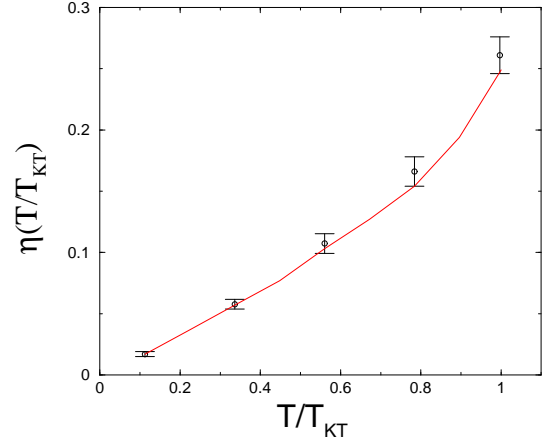}}
\end{center}
\caption{Dependence of the $\eta$ exponent on temperature extracted from the two-time autocorrelation function (symbols).
The solid line corresponds to numerical values given in ref.\cite{berche}. \label{fig2}
}
\end{figure}

In the asymptotic regime, the two-times response function is expected to be given, at least at low temperature by the spin-wave approximation
formula~(\ref{sec26}). In figure~3, we give the numerical results obtained for a final temperature of $T=0.3$ for different waiting times.
\begin{figure}[ht]
\epsfxsize=7cm
\begin{center}
\mbox{\epsfbox{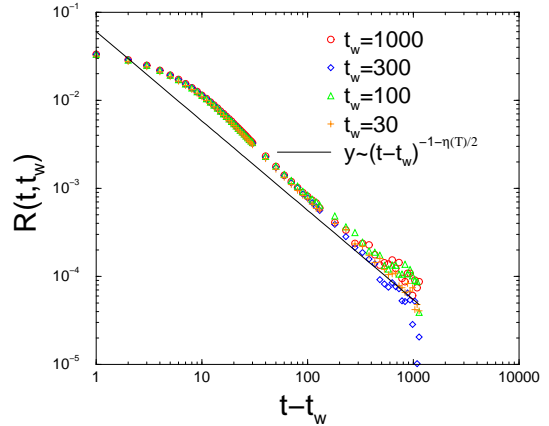}}
\end{center}
\caption{Response function at $T=0.3$ for different waiting times. \label{fig3}
}
\end{figure}
\begin{figure}[ht]
\epsfxsize=7cm
\begin{center}
\mbox{\epsfbox{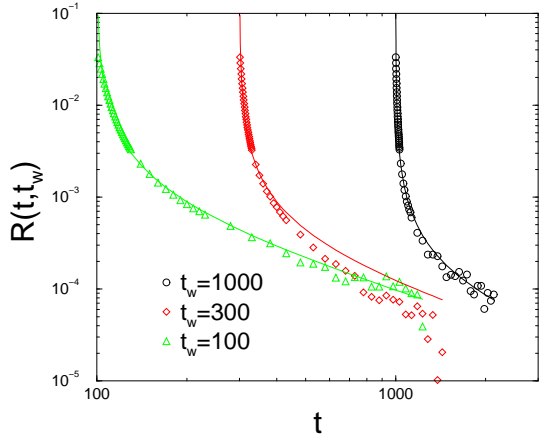}}
\end{center}
\caption{Response function for different waiting times for a system of linear size $L=512$. The quench temperature is $T=0.3$.
The solid lines correspond to the analytical expression (\ref{sec26}). \label{fig4}
}
\end{figure}
The aging part of the response is very small in the accessible regime and the deviation from a time-invariant process is very difficult to be seen
as the collapse of the data for different waiting times in figure~3 attests.
Nevertheless, we have plotted in figure~4 directly the numerical
response function together with the analytical prediction. The superposition of both curves seems to validate the expected law.
Although, in ref.\cite{berthier} this case was considered quite extensively, it was done only for
one temperature. Here we have extended the results to the whole low temperature regime.

\subsection{Infinite temperature initial state}
The infinite-temperature initial state is more canonical in the study of coarsening and aging effects. After the quench in the critical phase,
the correlation length will grow in time with a logarithmic correction due to the interaction of walls with free vortices as mentioned in ref.\cite{bray2}.
That is, $\xi(t)\sim(t/\ln t)^{1/2}$ and leading to the conjectures~(\ref{sec28},\ref{sec29}) for the correlation and response functions.
Berthier {\it et al} have checked this conjecture for the correlation length only for one final temperature, namely $T=0.3$. In figure~5 we show the
results obtained for several quench temperatures, ranging from $T=0.1$ up to $T=0.7$. The collapse of the rescaled correlation functions
$(t-t_w)^{\eta/2}C(t,t_w)$ as function of the variable $\xi(t)/\xi(t_w)$ is very satisfactory.  From these curves, we can extract the scaling
function $f_C$, see equation~(\ref{sec28}), and find the power law behaviour $f_C(x)\sim x^{-\kappa}$ with a temperature independent exponent
$\kappa=1.05(10)$. In ref.\cite{berthier} the value $\kappa = 1.08$ was given which is of course compatible with our data.
\begin{figure}[ht]
\epsfxsize=7cm
\begin{center}
\mbox{\epsfbox{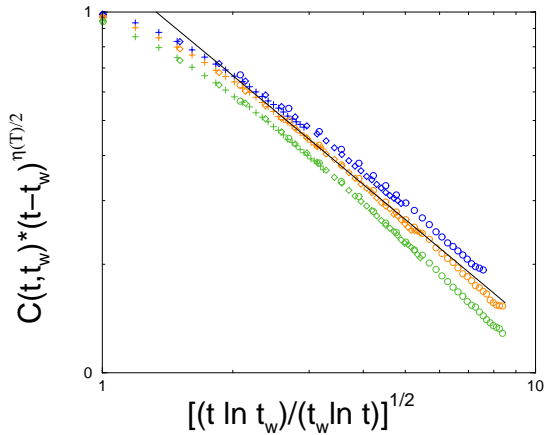}}
\end{center}
\caption{Scaling plot of the two-times autocorrelation function for a quench from infinite temperature towards $T<T_{KT}$.
The three different collapsed curves are obtained for $T=0.3$, $T=0.5$ and $T=0.7$ from top to bottom.
The different waiting times for each temperature are $t_w=100$ circles, $t_w=300$ diamonds and $t_w=1000$ crosses.
The solid line corresponds to $1/x$.\label{fig5}
}
\end{figure}
However, one has to be careful in this statement since the exponent is very close to $1$. For example, if one takes as the scaling variable
$x^{-1}=\xi(t_w)/\xi(t)$ instead of $x=\xi(t)/\xi(t_w)$, then the scaling limit we are interested in is $x^{-1}\ll 1$
and what is seen could be very possibly the leading expansion terms of an analytical scaling function, that is $g(x^{-1})\simeq g(0)+\alpha x^{-1}$.
Moreover, numerically the extrapolated value $g(0)$ seems to be very close to zero (less than $0.01$) and it is impossible to test this value in the time
range explored in this work. The same is true for the reference~\cite{berthier}. So if $g(0)$ is not vanishing, finally at long enough times, the decay
of the autocorrelation function has the same power law dependence as in the equilibrium situation. Otherwise, the decay is faster with
a power law $t^{-\eta/2-1/2}$ up to logarithmic corrections.

The parametric plot of the susceptibility times the temperature versus the correlation function does not collapse 
for different waiting times, showing
that the fluctuation-dissipation ratio is not a function of $C$ only but rather has a dependence on both $t$ and $t_w$.
However, after an initial quasi-equilibrium regime, where the different waiting time curves are collapsing and lead to the
equilibrium value $X(t,t_w)=1$, they have a constant slope, $X^{\chi}_{\infty}$, independent of $t_w$. This number
$X^{\chi}_{\infty}$ corresponds, when $X(t,t_w)=X(C(t,t_w))$ only, to the asymptotic limit
$X_{\infty}=\lim_{t_w\rightarrow\infty}\lim_{t\rightarrow\infty}X(t,t_w)$.
In figure~6, where we have represented $X^{\chi}_{\infty}$ versus the reduced temperature $T/T_{KT}$,
we clearly see a linear behaviour, starting at $X^{\chi}_{\infty}=0$ for $T=0$ and finishing at the value $X^{\chi}_{\infty}=1/2$
at the Kosterlitz-Thouless point.
It can be noticed that this continuous dependence of the FDT ratio on a model parameter
was also observed in the spherical model.\cite{picone}
\begin{figure}[ht]
\epsfxsize=7cm
\begin{center}
\mbox{\epsfbox{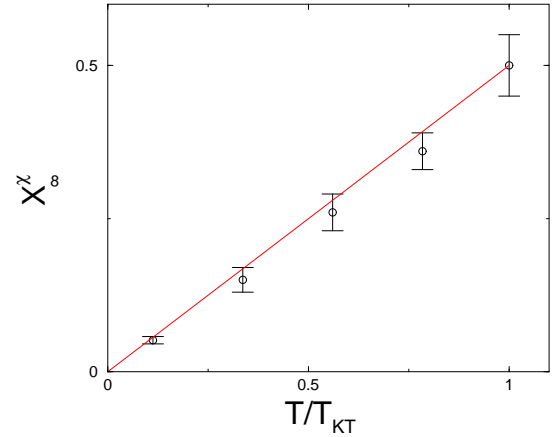}}
\end{center}
\caption{ Fluctuation-dissipation ratio versus reduced temperature $T/T_{KT}$. The line is the conjectured function.\label{fig6}
}
\end{figure}

\begin{figure}[ht]
\epsfxsize=7cm
\begin{center}
\mbox{\epsfbox{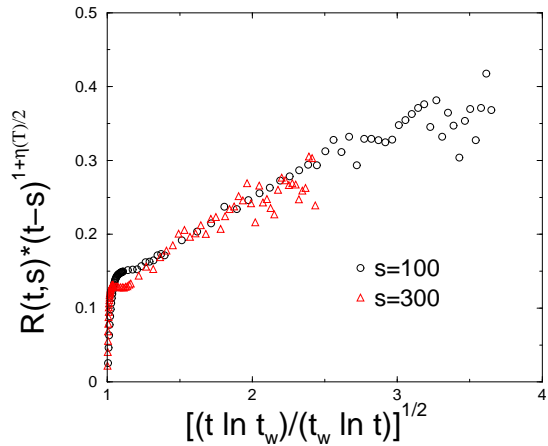}}
\end{center}
\caption{ Rescaled response function at $T=0.1$ for a linear system size $L=100$ averaged over 11000 realizations.\label{fig7}
}
\end{figure}

\begin{figure}[ht]
\epsfxsize=7cm
\begin{center}
\mbox{\epsfbox{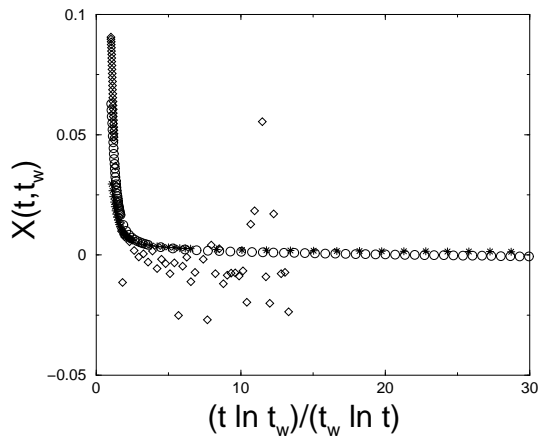}}
\end{center}
\caption{ Fluctuation-dissipation ratio at $T=0.1$ as a function of the scaling variable $t \ln t_w/t_w \ln t$
for $t_w=10$ (stars), $t_w=30$ (circles) and $t_w=100$ (diamonds).\label{fig8}
}
\end{figure}
Finally, we present the data obtained for the linear response function and the fluctuation-dissipation ratio $X(t,t_w)$.
The simulations are done on lattices of linear size up to $L=100$ averaged
over 11000 realizations in order to have a good enough statistics for $X$.
In figure~7, we show the result obtained for a final temperature $T=0.1$. Very similar curves are obtained for other temperatures.
The collapse of the data for different waiting times is very good, which confirms the scaling conjecture~(\ref{sec29}). Although the number of different
histories we have realized is quite huge, the noise on the points is still important. The range of time used is from $t=100$ up to $t=2500$, which
explains the very short window of the x-axis in figure~7. Nevertheless, what it is seen after an initial short time regime is a linear behaviour in terms of the scaling variable
$\xi(t)/\xi(t_w)$, that is asymptotically 
\begin{equation}
(t-t_w)^{1+\eta/2}R(t,t_w)\simeq A_R \left(\frac{t\ln t_w}{t_w\ln t}\right)^{1/2}\; ,
\end{equation}
with an amplitude $A_R$ slowly varying with the temperature.
In figure~8, we present the fluctuation-dissipation ratio $X(t,t_w)$ as a function of $t \ln t_w/t_w \ln t$
obtained numerically at the same temperature, $T=0.1$, for waiting times $t=10,30,100$.
Since in the calculation of $X$, we have to do a derivative, the results obtained are much more noisy and it is difficult to go on very long waiting times.
Nevertheless, we clearly see a good collapse on a master curve, leading to a vanishing fluctuation-dissipation ratio in the asymptotic limit. The same
is obtained for other temperatures.

\section{Summary and outlook}
We have studied numerically the non-equilibrium relaxation properties of the two-dimensional XY model initially prepared on two distinct ways:
completely ordered and fully disordered state. In both cases, the two-time spin autocorrelation function and the associated linear response
function were determined.

In the initial ordered case, we have fully confirmed the analytical predictions obtained in the spin wave approximation, strictly
valid at very low temperature. Nevertheless, the scaling form given in equation (\ref{sec25}) seems to be valid in a wide range of temperature
below the Kosterlitz-Thouless transition. From it, we have extracted the equilibrium exponent $\eta(T)$  with a quite good
accuracy as seen in figure~2.
Using the Glauber-like dynamics defined previously, we have obtained directly the linear response function itself. This permits a direct comparison
of our data with the analytical expression (\ref{sec26}).
With this approach, we avoided difficulties inherent in the use of susceptibilities, which can be affected by short-time contributions.
Although the aging part of the response seems to be very small, as attested in figure~3, the plot in figure~4 shows a very good agreement between the
numerical data and the analytical prediction.

Starting with a fully disordered state, we have extended the conjecture checked in ref.\cite{berthier} for one particular temperature
to the whole low temperature regime. For temperature ranging from $T=0.1$ up to $T=0.9$, we have numerically confirmed the forms~(\ref{sec28}) and
(\ref{sec29}) of the correlation and response functions with a scaling variable given in (\ref{sec30}).
As discussed previously, we found numerically for the asymptotic behaviour of the autocorrelation scaling function $f_C$, defined in (\ref{sec29}),
a behaviour which is compatible with a purely algebraic decay with a temperature independent exponent very close to one.
The linear response scaling function, in the time-range studied here, has a linear behaviour in the scaling variable $\xi(t)/\xi(t_w)$.
Those asymptotic behaviours of the correlation and response scaling functions are supporting, up to logarithmic factors, 
the forms given by local scale-invariance theory.\cite{henkel1}
Utilising the notations of \cite{godreche2}, one has
\begin{eqnarray}
C(t,t_w)\simeq t_w^{-a}F_C(t/t_w)\label{eqc}\\
R(t,t_w)\simeq t_w^{-a-1}F_R(t/t_w)\label{eqr}\; ,
\end{eqnarray}
where the scaling functions $F_C$ and $F_R$ have the asymptotic forms
\begin{equation}
F_{C,R}(u)\simeq A_{C,R}\; u^{-\lambda_{C,R}/z}\qquad u\gg 1\; .
\end{equation}
From our data, we obtain $a=\eta(T)/2$ and $\lambda_{C}=\lambda_{R}=\eta(T)+1$ confirming the general scenario depicted 
in ref.\cite{henkel1,godreche2,henkelpaessens}.
Finally, we extracted $X^{\chi}_{\infty}$, from the parametric susceptibility/correlation plot in the long-time limit.
The result we obtained is well fitted by the linear behaviour $X^{\chi}_{\infty}=(1/2) T/T_{KT}$.
The direct use of the response function gave a different answer, as seen in figure~8. Although the fluctuation-dissipation ratio $X(t,t_w)$ is a function
of both $t$ and $t_w$, this dependence seems to enter only through the scaling ratio $t\ln t_w/t_w\ln t$. From the numerical results we obtained on the
correlation and response functions, it is clear that in the asymptotic regime the fluctuation
dissipation ratio $X_{\infty}=\lim_{t_w\rightarrow\infty}\lim_{t\rightarrow\infty}X(t,t_w)$ is vanishing.
A result which is different from what is obtained from the parametric susceptibility/correlation plot.
This vanishing
is due to the breakdown of scaling induced by the presence of the logarithmic factors in the scaling functions.
In practice, one has to take care when discussing those plots, especially when no master curve is ever reached
for different waiting times.

\section*{Acknowledgments}
Thanks to C. Chatelain and M. Henkel for the support they offered to us. The other members of the Groupe de Physique Statistique are
greatfully acknowledged too.

\end{document}